\documentclass[final,3p,times,twocolumn]{elsarticle}

\usepackage{graphicx}
\usepackage{dcolumn}
\usepackage{bm}
\usepackage{amsmath}
\usepackage{amssymb}
\usepackage[utf8]{inputenc}
\usepackage[T1]{fontenc}
\usepackage{etoolbox}
\usepackage{hyperref}
\usepackage{xcolor}
\hypersetup{
    colorlinks,
    linkcolor={red!50!black},
    citecolor={blue!50!black},
    urlcolor={blue!80!black}
}

\begin{document}

\let\today\relax
\makeatletter
\def\ps@pprintTitle{%
    \let\@oddhead\@empty
    \let\@evenhead\@empty
    \def\@oddfoot{\footnotesize\itshape
         {} \hfill\today}%
    \let\@evenfoot\@oddfoot
    }
\makeatother


\begin{frontmatter}

\title{Axisymmetric Coil Winding Surfaces for Non-Axisymmetric Fusion Devices}



\author[IST]{J. Biu}
\author[IST,UW-M]{R. Jorge}
\affiliation[IST]{organization={Instituto de Plasmas e Fusão Nuclear, Instituto Superior Técnico, Universidade de Lisboa, 1049-001 Lisboa, Portugal}}
\affiliation[UW-M]{organization={Department of Physics, University of Wisconsin-Madison, Madison, Wisconsin 53706, USA}}

\begin{abstract}
Stellarators are fusion energy devices that confine a plasma using non-axisymmetric magnetic fields. Complex coils with tight construction tolerances are needed to create such fields. To simplify such coils, we use a method here to create filamentary curves bounded to a coil winding surface. This approach bypasses the need to find contours of the current potential in that surface while allowing gradients to be obtained for both the winding surface and the coil shapes. The parameterization of the coil curves allows the modeling of both modular and helical coils. As an application, we optimize a set of coils to reproduce a quasisymmetric stellarator equilibrium. A comparison is performed between coils parameterized in two different winding surfaces, namely an axisymmetric circular toroidal surface and a surface rescaled from the plasma boundary. Finally, an analysis is performed on the optimal distance between the plasma and the coil winding surface.
\end{abstract}

\end{frontmatter}

\section{\label{sec:introduction}Introduction}

Magnetic confinement fusion energy relies on coils placed carefully around a plasma chamber to confine a burning plasma. The coils are typically one of the most expensive components of the machine, especially in non-axisymmetric geometries.
Tokamaks and stellarators are two types of toroidal magnetic confinement fusion devices that wrap field lines around magnetic flux surfaces \cite{Freidberg2007}. Tokamaks are axisymmetric and have a toroidal magnetic field created by external coils and a poloidal magnetic field created by inducing a toroidal plasma current, resulting in field lines that are toroidal helices. On the other hand, stellarators do not need an externally driven plasma current, as their poloidal field is mainly produced by the external coils. Therefore, they can run in a steady state \cite{Helander2014}. However, the breaking of axisymmetry, caused by having a non-planar magnetic axis, or by making the flux surfaces rotate poloidally along the toroidal angle \cite{Xu2016}, results in increased coil complexity.  
This has proven to be a great challenge for building stellarators since such coils are usually expensive, difficult to design and build, and have tight tolerances \cite{Neilson2010LessonsNCSX}. Therefore, the geometry of the magnetic flux surfaces, as well as the geometry of the coils needed to create them, is crucial for making these devices viable. 

Many optimization tools and methods have been developed to obtain configurations for the geometry of the coils capable of creating magnetic flux surfaces that confine a plasma while minimizing cost and construction difficulties.
The optimization of stellarators is usually divided into two stages, namely the optimization of the boundary magnetic flux surface and the optimization of the coils. This work focuses on the second stage, coil optimization. While several single-stage approaches have been developed recently \cite{Giuliani2022,Jorge2023}, this can be done independently of the boundary of the magnetic field and can take into account many parameters to regularize the shape of the coils  \cite{Lobsien2017,Hudson2018}. For this purpose, several optimization codes have been developed, such as NESCOIL \cite{Merkel1987} and REGCOIL \cite{Landreman2017}, FOCUS \cite{Zhu2017a}, or SIMSOPT \cite{Landreman2021b}.

NESCOIL obtains the stellarator coils by optimizing a continuous current potential surface on the coil-winding surface (CWS) where the coils lie. This code searches for the current potential that produces the target plasma shape, and once this is obtained, the discrete coils result from taking closed contours of the coil winding surface. REGCOIL is a variant of NESCOIL that adds average squared current density as an optimization criterion to the code, leading to coils that are regularized.
On the other hand, FOCUS presents a different approach to the optimization of coils. Instead of starting with a current density on a surface that is later divided into coils, FOCUS optimizes discrete coils by representing them as closed one-dimensional curves in space. The optimization of these objects is then done by minimizing an objective function where the targets, constraints, and thresholds are defined. This approach is also used in SIMSOPT. This direct optimization of the coils, instead of optimization of a whole surface, was shown to be viable to obtain coils for already existing stellarators such as W7-X \cite{Drevlak1998}.

The tool used in this work is SIMSOPT.
This is a framework for stellarator optimization that is able to optimize both of the boundary flux surface and its corresponding magnetic field inside (stage one) and the external coils (stage two) using a FOCUS approach that reproduces the target magnetic field obtained in stage one. In SIMSOPT, coils are defined as parameterized curves with coordinates given by a Fourier series, where the degrees of freedom for the optimization are the Fourier coefficients and the current applied to each coil \cite{Wechsung2022}. The optimization searches for the best configuration of coils by comparing the magnetic field that they create with the target flux boundary surface. The method employed here relies on such a Fourier decomposition. However, instead of parameterizing the Cartesian components $(x,y,z)$ using a curve parameter $t$, we parameterize the plasma boundary surface using its cylindrical components $(R,Z)$ using a curve parameter $t$.

In this work, coil winding surfaces are set as fixed objects where we create space curves parameterized to them, which are then optimized while bound to the surface. When applied to designs with constrained winding surfaces, such as coil design methods using deposition on a surface, this allows us to be more numerically efficient than the standard FOCUS approach, as it decreases the number of free parameters and optimization constraints. Furthermore, shape optimization for the winding surface can be performed in a single-stage manner, as derivatives of the surface shape are readily available.
As shown in Ref. \cite{Queral2013}, methods based on coil winding surfaces, such as milling and direct deposition, are able to avoid expensive metrology systems as no adjustments are needed for modular coils or frames, and no winding moulds and fasteners are needed.
Using the method outlined here, we obtain a set of coils bound to a given CWS that produce a quasisymmetric stellarator. Furthermore, a study is performed on the optimal surface-to-coil distance for a coil winding surface that is obtained either by expanding the plasma boundary outward or using a circular torus as a CWS. Such a simple surface is shown to be a viable winding surface for the coils.

This paper is organized as follows. In Section~\ref{sec:opt}, the optimization methods and tools for the design of stellarators will be further explored, as well as the codes used, and in Section~\ref{sec:implementation}, the implementation of our approach is described. Finally, the results obtained are shown and analyzed in Section~\ref{sec:results}. The conclusion follows.

\section{\label{sec:opt}Optimization Tools and Methods}

To confine the plasma, the first step is to obtain a suitable magnetic field equilibrium.
This can be achieved by solving the static ideal magnetohydrodynamics (MHD) system of equations, which considers the plasma to be a single, electrically conducting fluid without resistivity. Ideal MHD relates the magnetic field $\mathbf{B}$, with the plasma current $\mathbf{J}$ density and the plasma pressure gradient $\nabla \textit{p}$ in a single equation known as the force balance equation
\begin{equation}
    \mathbf{J} \times \mathbf{B} = \nabla p .
    \label{eq:ideal_mhd}
\end{equation}
By solving Eq.~(\ref{eq:ideal_mhd}), we can obtain a magnetic field that confines the plasma and obeys Maxwell's equations, namely 
$\nabla \times \mathbf{B} = \mu_0 \mathbf{J}$ and $\nabla \cdot \mathbf{B} = 0.$
Ideal MHD is valid assuming a high collisionality regime and characteristic frequencies lower than the electron and ion gyro frequencies \cite{Freidberg2014}. 
An important property of the force balance equation, Eq.~(\ref{eq:ideal_mhd}), is that the magnetic field $\mathbf{B}$ and current $\mathbf{J}$ vectors are coincident with constant pressure surfaces, as
\begin{equation}
\mathbf{B} \cdot \nabla p = \mathbf{J} \cdot \nabla p = 0,
\end{equation}
showing that flux surfaces have constant pressure.
Poicaré Index Theorem then implies that a magnetic field that confines a plasma must have a toroidal topology \cite{Helander2014}. The necessary magnetic field for plasma confinement is one with flux surfaces nested inside each other.
When optimizing for coils, we take advantage of this property and use the last closed flux surface of constant plasma pressure (plasma boundary) as a target surface for the coil magnetic field. Therefore, our goal is to make the coil magnetic field parallel to the surfaces of constant pressure, $\mathbf B \cdot \mathbf n = 0$ where $\mathbf n$ is the vector normal to the surface.

In this work, the magnetic field $\mathbf{B}$ is found using VMEC (Variational Moments Equilibrium Code) \cite{Hirshman1983}, which solves the MHD equations assuming a toroidal domain for the equilibrium solution and nested flux surfaces. VMEC obtains the MHD equilibrium by minimizing the potential energy $W = \int [{B^2}/{(2\mu_0)} + {p}/{(\gamma - 1)}]d^3x$ of the plasma with $\gamma$ the adiabatic index using the steepest descent method. We use VMEC in a fixed boundary mode, where the magnetic field $\mathbf B$ is found using force balance and the last closed flux surface as a boundary condition.
The plasma boundary is parameterized using a double Fourier series in cylindrical coordinates, $ (R, Z, \phi)$, in the form
\begin{equation}
    \label{eq:surface_R}
    R(\theta, \phi) = \sum_{m = 0}^{m_{\text{pol}}}\sum_{n = -n_{\text{tor}}}^{n_{\text{tor}}} \text{RBC}_{m, n} \cos{(m\theta - n_{\text{fp}} n \phi)},
\end{equation}
and
\begin{equation}
    \label{eq:surface_Z}
    Z(\theta, \phi) = \sum_{m = 0}^{m_{\text{pol}}}\sum_{n = -n_{\text{tor}}}^{n_{\text{tor}}} \text{ZBS}_{m, n} \sin{(m\theta - n_{\text{fp}} n \phi)},
\end{equation}
where $\theta \in [0, 2\pi[$ is a poloidal angle, $\phi \in [0, 2\pi[$ is the standard cylindrical angle and $n_{\text{fp}}$ is the number of field periods. The Fourier coefficients are given by $\text{RBC}_{m, n}$ and $\text{ZBS}_{m, n}$ and since we are working with surfaces that are symmetric under rotation by $\pi$ about the x-axis, i.e., have stellarator-symmetry, the sine terms for $R$ and the cosine terms for $Z$ are zero. The position vector $\mathbf{r}$ for the surface is then given in cylindrical coordinates as $\mathbf{r}$ = $[(R(\theta, \phi)\cos{(\phi)},\; R(\theta, \phi)\sin{(\phi)},\; Z(\theta, \phi)]$, and the degrees of freedom of the surface are then
\begin{equation}\label{eq:xsurface}
    \mathbf{x}_{\text{surface}} = [\text{RBC}_{m, n}, \text{ZBS}_{m, n}].
\end{equation}

Analogous to the way that surfaces are parameterized using two parameters ($\theta$, $\phi$), the external coils that create the magnetic field can be considered as closed curves parameterized by a single parameter $t$. 
A typical approach to model filamentary coils, as it is used in FOCUS and SIMSOPT, is to write the coordinates $\mathbf{r}=(x,y,z)$ of the curve as a truncated Fourier series
\begin{equation}
    \label{eq:curve_x}
    x(t) = \sum_{m=0}^{\text{order}} x_{c,m}\cos{(m~t)} + \sum_{m=1}^{\text{order}} x_{s,m}\sin{(m~t)},
\end{equation} 
with $t \in [0, 2\pi[$, $\text{order}$ the order of the Fourier expansion and $m$ integers. The parameterization for $y(t)$ and $z(t)$ are similar to Eq.~(\ref{eq:curve_x}). This way, to create the coils, the degrees of freedom are the Fourier coefficients
\begin{equation}
\mathbf{x}_{\text{coils}} = [x_{c,m}, x_{s,m}, y_{c,m}, y_{s, m}, z_{c, m},  z_{s, m}].   
\end{equation}
Given the fact that both the coils and surfaces are written as Fourier series, which are continuous and differentiable, their derivatives can be computed analytically, as opposed to relying on numerical methods such as finite differences. 

Such an approach has been shown to allow for coils that accurately reproduce quasisymmetric fields \cite{Wechsung2022} using coil optimization.
When performed in vacuum fields, the optimization relies on the shape of the plasma boundary only and targets a magnetic field with field lines aligned with that surface.
This is done by minimizing an objective function $J$, defined in Eq.~(\ref{eq:quadratic_flux}), entitled the quadratic flux, which is the surface integral of the square of the normalized error field, i.e., the magnetic field created by the coils normal to the surface divided by the total field
\begin{equation}
    {J} =\frac{1}{2} \int \frac{|\mathbf{B}_{\text{coils}}\cdot \mathbf{n}|^2}{|\mathbf{B}_{\text{coils}}|^2} d^2x,
    \label{eq:quadratic_flux}
\end{equation}
where $\mathbf{n}$ is the normal vector to the boundary surface at each grid point. The magnetic field created by the coils is denoted as $\mathbf{B}_{\text{coils}}$ and is computed using the Biot-Savart law
\begin{equation}
    \mathbf B_{\text{coils}}({\mathbf r})=\frac{\mu_0}{4\pi}\sum_{i=1}^{2n_{\text{fp}}N_C} I_i \int_{\mathbf \Gamma_i}\frac{d \mathbf l_i\times \mathbf r'}{r'^3},
\end{equation}
$I_i$ is the current in each coil, where $d \mathbf l_i = \mathbf r_i dt$ is the differential line element of the coil $i$-th curve $\mathbf{r}_{\text{coil}}$ and $\mathbf r'={\mathbf r} - \mathbf r_i$ is the displacement vector between the evaluated point on the surface and the differential element.
Each coil is divided into an integer number of quadrature points, typically between 100 and 200, and the minimization is performed using the gradient-based L-BFGS-B method \cite{Liu1989}, which requires cost function gradients.
Their implementation is detailed in the next section.

While optimizing coil shapes to yield $J=0$ would lead to coils that accurately reproduce the target magnetic field, such a minimization problem is ill-posed \cite{Landreman2017,Zhu2017a}.
Therefore, the space of allowable coil shapes is restricted by penalizing coil complexity using the method introduced in Ref. \cite{Wechsung2022}, leading to simpler coils. We consider the following regularization terms
\begin{align}
    g_L&=\varphi\left(\sum_{i=1}^{N_c}L_i-L_{\text{max}}\right),\\
    g_{\kappa,\text{max}}&=\sum_{i=1}^{N_c}\int_0^{2\pi}\mathrm{M}(\kappa_i-\kappa_{\text{max}})|\Gamma^{'(i)}|d\theta/L_i,\\
    g_{\kappa,\text{msc}}&=\sum_{i=1}^{N_c}\varphi\left(\int_0^{2\pi}\kappa_i^2|\Gamma^{'(i)}|d\theta/L_i-\kappa_{\text{msc}}\right),\\
    g_d&=\sum_{i=1}^{N_C}\sum_{j=1}^{i-1}\int_0^{2\pi}\int_0^{2\pi}\mathrm{M}(d_{\text{min}}\nonumber\\
    &-|\Gamma^{(i)}(\theta)-\Gamma^{(j)}(\theta')|)|\Gamma^{'(i)}(\theta)\Gamma^{'(j)}(\theta')|d \theta d \theta',
\end{align}
where $L_i$ and $\kappa_i$ are the length and curvature of the coil $i$, respectively, and $\mathrm{M}(t)=\text{max}(t,0)^2$ is the differentiable maximum function.
The term $g_d$ intends to maximize the distance between coils for maximum port accessibility.
These allow us to restrict the total length of the independent coils to be $L_{\text{max}}$, the individual coil curvature and mean squared curvature to $\kappa_{\text{max}}$ and $\kappa_{\text{msc}}$, respectively, and the minimum distance between coils to $d_{\text{min}}$.
Such regularization terms are employed in order to restrict the space of allowed coil shapes to satisfy the following set of conditions
\begin{subequations} \label{eq:opt_conds}
\begin{align}
    \sum_{i=1}^{N_c}L_{c}^{(i)}&\le L_{\text{max}},\\
    \kappa_i &\le \kappa_{\text{max}},~i=1,\ldots,N_c,\\
    \frac{1}{L_c^{(i)}}\int_{\gamma^{(i)}}\kappa_i^2 dl &\le \kappa_{\text{msc}},~i=1,\ldots,N_c,\\
    \| \Gamma^{(i)}-\Gamma^{(j)} \| &\geq d_{\min} ~ \text{ for } i \neq j,\label{eq:optprob:8}.
\end{align}
\end{subequations}
Each cost function $g$ is added to the square flux $J$ to obtain the total loss used in the optimization.

\section{\label{sec:implementation}Implementation of Surface-Bounded Coils}

To implement the direct optimization of filamentary coils in a coil winding surface, we start by defining the parameterization of the curve on the surface. Given the definition of a toroidal surface in Eqs.~(\ref{eq:surface_R}) and (\ref{eq:surface_Z}), our task is to parameterize $\theta$ and $\phi$ using a single parameter $t$. 
We define $\theta$ and $\phi$ as a sum of secular terms in $t$ together with a Fourier series. This is written as
\begin{equation}
    \label{eq:curve_cws_theta}
    \theta(t) = \theta_l \, t + \sum_{k=0}^{\text{order}} \theta_{c, k} \cos(k\,t) + \theta_{s, k} \sin(k\,t),
\end{equation}
and
\begin{equation}
    \label{eq:curve_cws_phi}
    \phi(t) = \phi_l \, t + \sum_{k=0}^{\text{order}} \phi_{c, k} \cos(k\,t) + \phi_{s, k} \sin(k\,t),
\end{equation}
where, as before, $\text{order}$ is the order of the Fourier series and $t \in [0, 2\pi]$.
Therefore, $\theta(t)$ and $\phi(t)$ are defined by a set of coefficients comprised by the linear terms, $\theta_l$, $\phi_l$, and the Fourier coefficients, leading to the following degrees of freedom
\begin{equation}
    \mathbf{x}_{\text{coils}} = [\theta_l, \; \theta_{c,k}, \; \theta_{s, k}, \; \phi_l, \; \phi_{c, k}, \; \phi_{s, k}],
    \label{eq:curvecwsfourier_dofs}
\end{equation}
together with the value of the current in each coil.
The expressions for $\theta(t)$ and $\phi(t)$ in Eqs.~(\ref{eq:curve_cws_theta}) and (\ref{eq:curve_cws_phi}) are then used as an input for $R$ and $Z$, Eqs.~(\ref{eq:surface_R}) and (\ref{eq:surface_Z}), so that the coordinates of a parameterized curve in a surface are given by
\begin{align}   
    \mathbf{r}(t) =& \left[R\left[\theta(t), \phi(t)\right]\cos{[\phi(t)]}, \right. \nonumber \\ & \left. R\left[\theta(t), \phi(t)\right]\sin{[\phi(t)]}, \; Z\left[\theta(t), \phi(t)\right]\right].
\end{align}
The parameterizations described by of Eqs.~(\ref{eq:curve_cws_theta}) and (\ref{eq:curve_cws_phi}) were implemented in SIMSOPT which takes as inputs the order of the Fourier series, the number of quadrature points to compute the curve, the degrees of freedom of $\theta(t)$ and $\phi(t)$ in Eq.~(\ref{eq:curvecwsfourier_dofs}), $\mathbf{x}_{\text{coils}}$, the currents, and the Fourier coefficients $\mathbf x_{\text{surface}}$ that define the winding surface in Eq.~(\ref{eq:xsurface}).

\begin{figure}
    \centering
    \includegraphics[trim={0.0cm 3cm 0.0cm 4cm}, clip, width=0.85\linewidth]{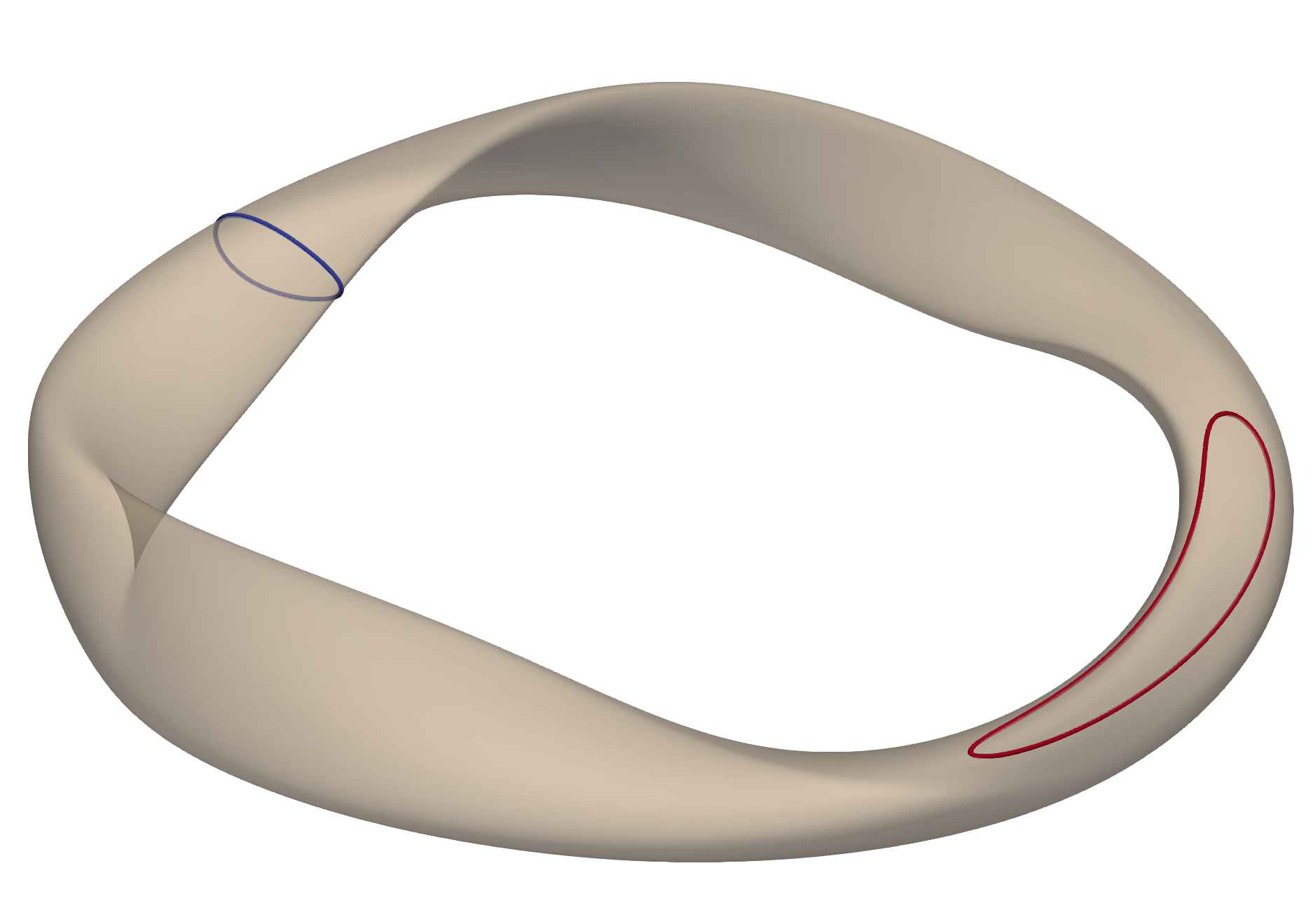}
    \caption{Example of curves parameterized to a surface with order=1. The parameters used to create them, according to Eq.~(\ref{eq:curvecwsfourier_dofs}), are $\mathbf{x}_{\text{coils}} = [1, \;0, \;0, \;0, \;0, \;0, \;0, \;0]$ for the poloidally closed curve in blue, and for the window pane curve in red, $\mathbf{x}_{\text{coils}} = [0, \;\pi/2, \;0.7, \;0, \;0, \;\pi, \;0, \;0.5]$.}
    \label{fig:curves1}
\end{figure}

\begin{figure}
    \centering
    \includegraphics[trim={0.0cm 3.5cm 0.0cm 7cm}, clip, width=0.85\linewidth]{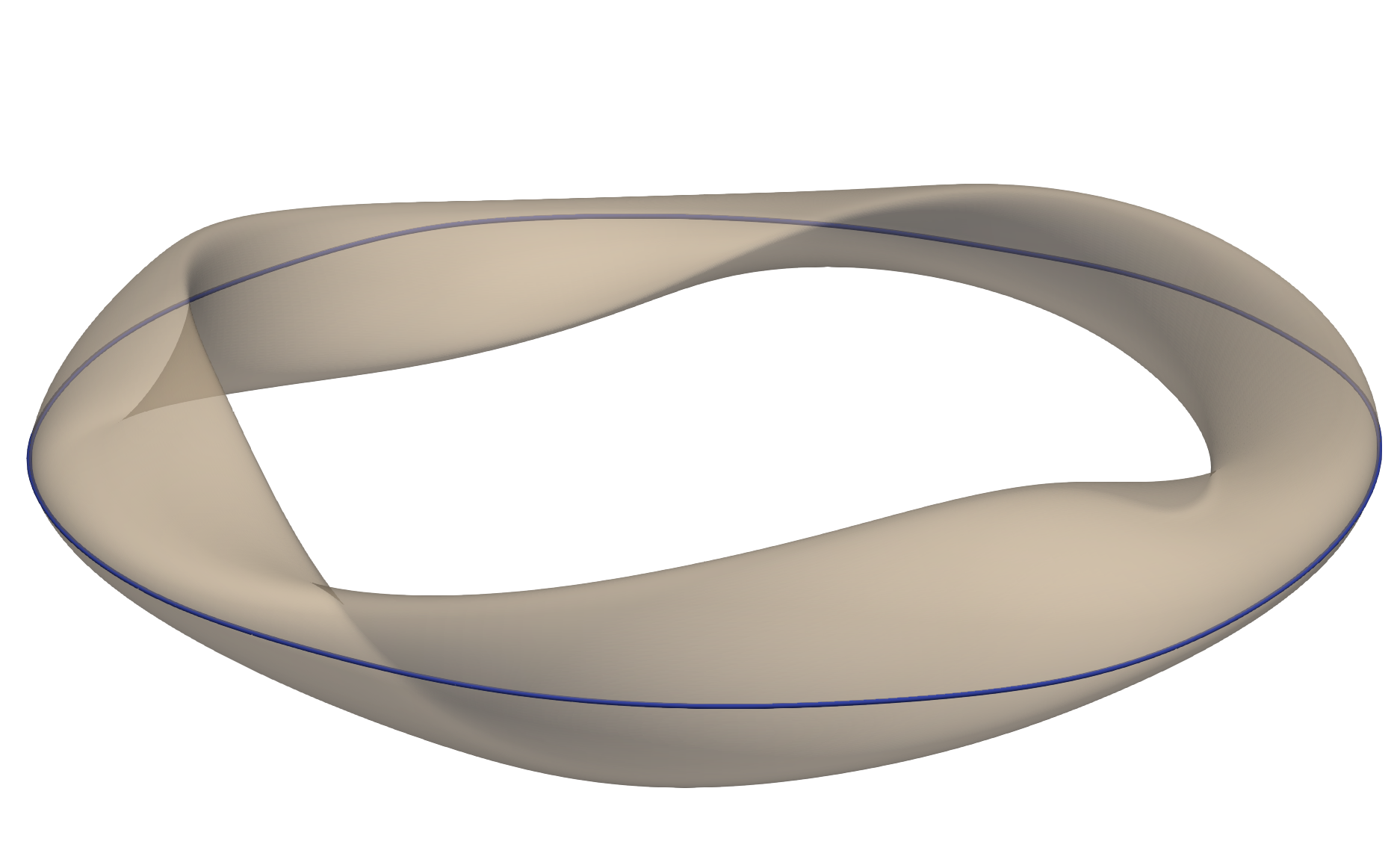}
    \caption{Example of toroidally closed curve parameterized to a surface with order=1. The parameters used to create it are $\mathbf{x}_{\text{coils}} = [0, \;0, \;0, \;0, \;1, \;0, \;0, \;0]$, according to Eq.~(\ref{eq:curvecwsfourier_dofs}).}
    \label{fig:curves2}
\end{figure}

In Figs.~\ref{fig:curves1} and \ref{fig:curves2} we show two examples of the new curves class using a non-axisymmetric CWS. These curves are created by recomputing the surface from its Fourier coefficients, and using Eqs.~(\ref{eq:curve_cws_theta}) and (\ref{eq:curve_cws_phi}) to find $\theta$ and $\phi,$ guaranteeing that the curves are bounded to the surface used as CWS. 

In order to perform gradient-based optimization involving the squared flux and curve quantities such as curvature, the first, second, and third derivatives of the curve coordinates, $\mathbf{r}($t$)$, with respect to the parameter $t$, were implemented, as well as the gradient of the coordinates with respect to the degrees of freedom defined in Eq.~(\ref{eq:curvecwsfourier_dofs}). The derivatives can be found analytically since the coordinates are defined in terms of a truncated Fourier series. As an example, the first derivative of the \textit{x}($t$) with respect to $t$ is given by
\begin{align}
\frac{dx(t)}{dt} &=
\sum_{m=0}^{m_{\text{pol}}}\sum_{n=0}^{2n_{\text{tor}}} \biggl[ -\text{RBC}_{m,\, n-n_{\text{tor}}} \sin\bigl(m\theta(t) - n_{\text{fp}} n \phi(t)\bigr)\biggr. \nonumber\\
& \times  \left(m\frac{d\theta(t)}{dt} - n_{\text{fp}} n \frac{d\phi(t)}{dt}\right)\cos(\phi(t))- \text{RBC}_{m,\, n-n_{\text{tor}}}\nonumber \\
& \left.\times\cos\left(m\theta(t) - n_{\text{fp}}n\phi(t)\right)  \sin\bigl(\phi(t)\bigr) \frac{d\phi(t)}{dt} \right].
\label{eq:derivative}
\end{align}

\begin{figure}
    \centering
    \includegraphics[width=0.85\linewidth]{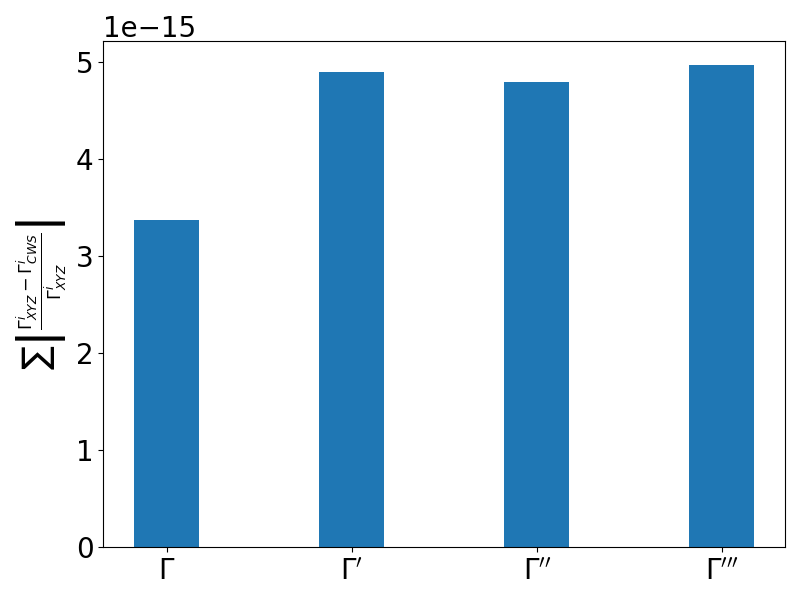}
    \caption{Relative error between the new coil curve implementation, CWS, and the coil curve class XYZ implemented in SIMSOPT used in previous coil optimization studies. From left to right, the horizontal axis labels the error in the coil positions, $\Gamma$, and their first, second, and third derivatives, respectively.}
    \label{fig:circular_CWS}
\end{figure}

In order to validate our implementation, we show in Fig.~\ref{fig:circular_CWS} a comparison between the new SIMSOPT curve class, entitled \textit{CWSFourier}, and a circular curve created using the \text{XYZFourier} class used in previous coil studies \cite{Wechsung2022,Jorge2023,Jorge2024}.
We use a circular torus with major radius of $R=1$ m and a minor radius of $r=0.1$ m as a coil winding surface.
Each curve has a total of 50 quadrature points, with the Fourier modes of the \textit{CWSFourier} coil being $\theta_l=1$ and order=0, while the Fourier modes of the \textit{XYZFourier} coil are $[x_{c0}=1,x_{c1}=0.1,z_{s1}=0.1]$ and order=1.
While this is a simple test case, it shows that such an implementation can lead to large savings in the number of degrees of freedom used in coil optimization.
We compute the coil positions, denoted in Fig.~\ref{fig:circular_CWS} as $\Gamma$, as well as their first, second, and third order derivatives with respect to their parameterization angle $t$, denoted as $\Gamma'$, $\Gamma''$, and $\Gamma'''$, respectively.
We find that the relative error between the two implementations, when summed over every point along the curve, is of the order of machine precision, $\sim 10^{-15}$, as expected.
We note that the magnetic field, as well as the coil simplicity metrics used here, are computed using such quantities.

\section{\label{sec:results}Optimization Results}

We now perform a coil optimization study to find the filamentary coils bound to a given CWS.
For this purpose, the boundary surface used was obtained from Ref.~ \cite{Jorge2024}, Section VI A, with a major radius of $1$ m and an aspect ratio of 10. This configuration was used since, due to the single-stage approach used there, the boundary is already optimized to take into account that coils lie on a winding surface. Indeed, such a boundary was obtained using helical coils bound to a circular torus, and has an approximately constant rotational transform of $\iota=0.41$ throughout the volume and has a positive magnetic well. While this is advantageous for this work, such a constraint leads to quasisymmetry being obtained only approximately. In the future, we plan to obtain precise quasisymmetric configurations using the CWS class developed here, which yields coils that can reproduce target fields with fewer constraints with respect to helical coils. The surface has five field periods, $n_{\text{fp}} = 5$, and we use 4 coils per field period and order=10 in Eqs.~(\ref{eq:curve_cws_theta}) and (~\ref{eq:curve_cws_phi}). We optimize coils for the lowest squared flux $J$ until a given tolerance in the optimization residual is achieved or for a given number of iterations. The following coil regularization constraints were used: coil length $L_{max}=20$ m, coil-to-coil distance $d_{min} = 0.1$ m, coil curvature $\kappa_{max}=60$ m$^{-1}$, and mean squared curvature $\kappa_{msc}=60$ m$^{-2}$.

\begin{figure}
    \centering
    \includegraphics[width=1\linewidth]{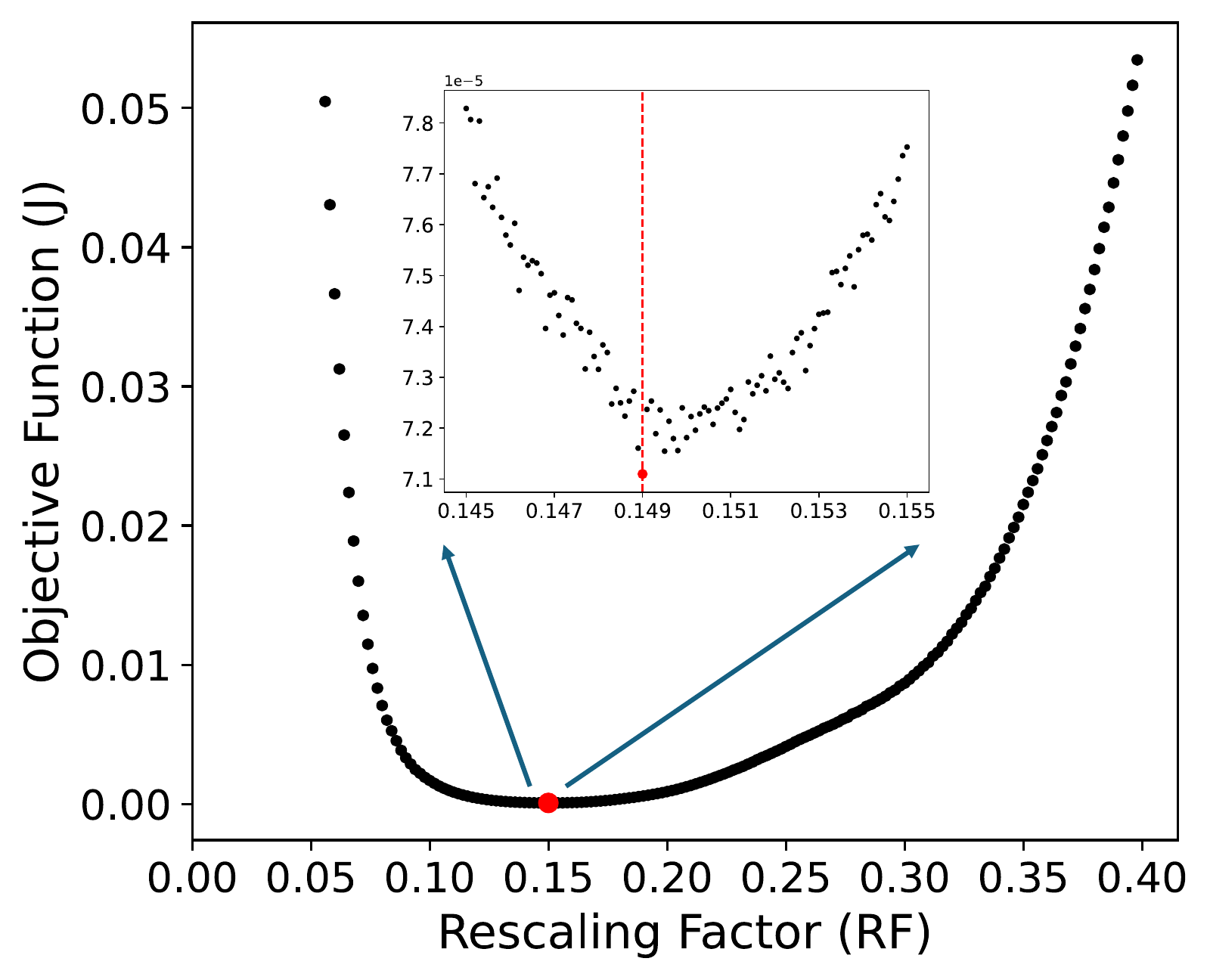}
    \caption{Variation of the objective function $J$ with the distance between the plasma boundary and CWS (RF) for the circular torus CWS. A scan is performed between 0.01 m and 0.4 m with a step value of $2\times 10^{-3}$ m. Then, a scan is performed between 0.145 m and 0.155 m with a step value of $1\times10^{-4}$ m. The value of RF obtained in this scan that minimizes the objective function is 0.149 m.}
    \label{fig:sweeping_circular}
\end{figure}

We first optimize a set of filamentary coils bound to a circular torus as the coil winding surface. To find the optimal dimension of the CWS and its distance to the plasma, we perform a series of coil optimizations with a maximum of 50 iterations for different dimensions of the CWS.
While a large plasma-coil distance is a requirement for a fusion power plant in order to have a shielding and blanket system, such large distances are known to degrade the ability of the coils to reproduce the target field and may lead to more complex coils.
The resulting squared flux $J$ from the set of optimizations is shown in Fig.~\ref{fig:sweeping_circular}. The dimension of the CWS is controlled by the distance from the major radius of the boundary surface to the major radius of the surface used as CWS, which we will address as the rescaling factor (RF). To obtain the value of RF that minimizes $J$, we perform two scans. The first in the interval from $0.01$ m to $0.40$ m with a step of $2\times 10^{-3}$ m is performed to find the approximate location of the global minimum of $J$. Once that minimum is found, around $0.15$ m, a scan is performed in the interval from $0.145$ m to $0.155$ m with a step of $1\times 10^{-4}$ m. The obtained value of RF that minimizes the squared flux J and, therefore, better reproduces the target magnetic field, is RF $ = 0.149$ m. This yields a CWS with a minor radius of 0.2565 m.

\begin{figure}[ht]
    \centering
    \includegraphics[trim={0.6cm 0.7cm 0.8cm 0.8cm}, clip, width=0.9\linewidth]{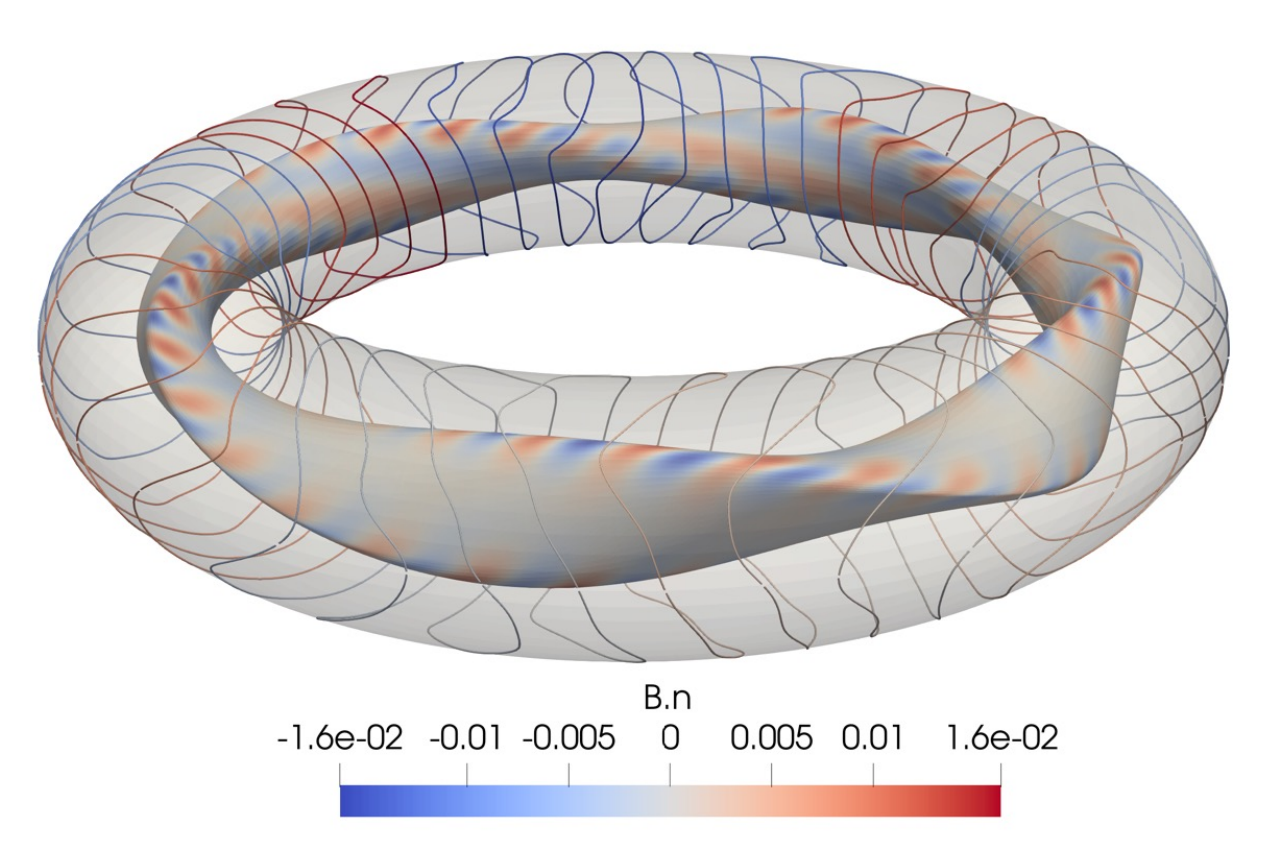}
    \caption{Set of coils obtained from an optimization in a coil winding surface with a circular toroidal shape targeting an approximately quasisymmetric magnetic field. The winding surface has a minor radius of 0.2565, and the distance between the plasma and the coils is 0.149 m. Both the winding surface and the plasma have a major radius of 1 m.}
    \label{fig:circular}
\end{figure}

\begin{figure}[ht]
    \centering
    \includegraphics[trim={0.0cm 0.40cm 0.0cm 0.0cm}, clip, width=1\linewidth]{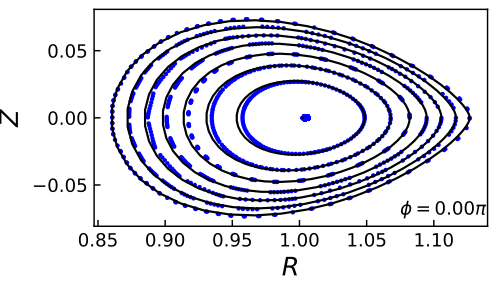}
    \caption{Superposition of magnetic surfaces of the target equilibrium at constant toroidal angle $\phi$ (black), and the Poincaré plot resulting from tracing magnetic field lines created by the optimized bounded coils (blue) in a circular toroidal CWS, at the cylindrical angle $\phi = 0$.}
    \label{fig:circular_poincare}
\end{figure}

Once the coil-plasma distance is obtained, a coil optimization using 2000 iterations of the L-BFGS-B algorithm is performed in order to minimize the objective function $J$.
The resulting coils are shown in Fig.~\ref{fig:circular}. For this set of coils, we find a value for the squared flux of $J$ in (Eq.~(\ref{eq:quadratic_flux})) of $5.316\times 10^{-5}$. The maximum field error on the surface is $\mathbf B \cdot \mathbf n = 1.6 \times 10^{-2}$ T with the target magnetic field being $B=1$ T.
To evaluate the effectiveness of the optimization using a CWS, we trace the magnetic field lines from the resulting Biot-Savart field of the coils and show a Poincaré section at cylindrical angle $\phi=0$ in Fig.~\ref{fig:circular_poincare}, superimposed with the target plasma boundary and interior flux surfaces obtained with VMEC. We note that Poincaré sections are formed by the intersection of the field line trajectories with a plane at a given cylindrical toroidal angle.
Here, we find that the field lines (blue) are able to produce a set of nested flux surfaces that closely match the target magnetic field surfaces computed using VMEC (black) for most of the domain.

\begin{figure}[ht]
    \centering
    \includegraphics[trim={0.0cm 0cm 0.0cm 0.0cm}, clip, width=1\linewidth]{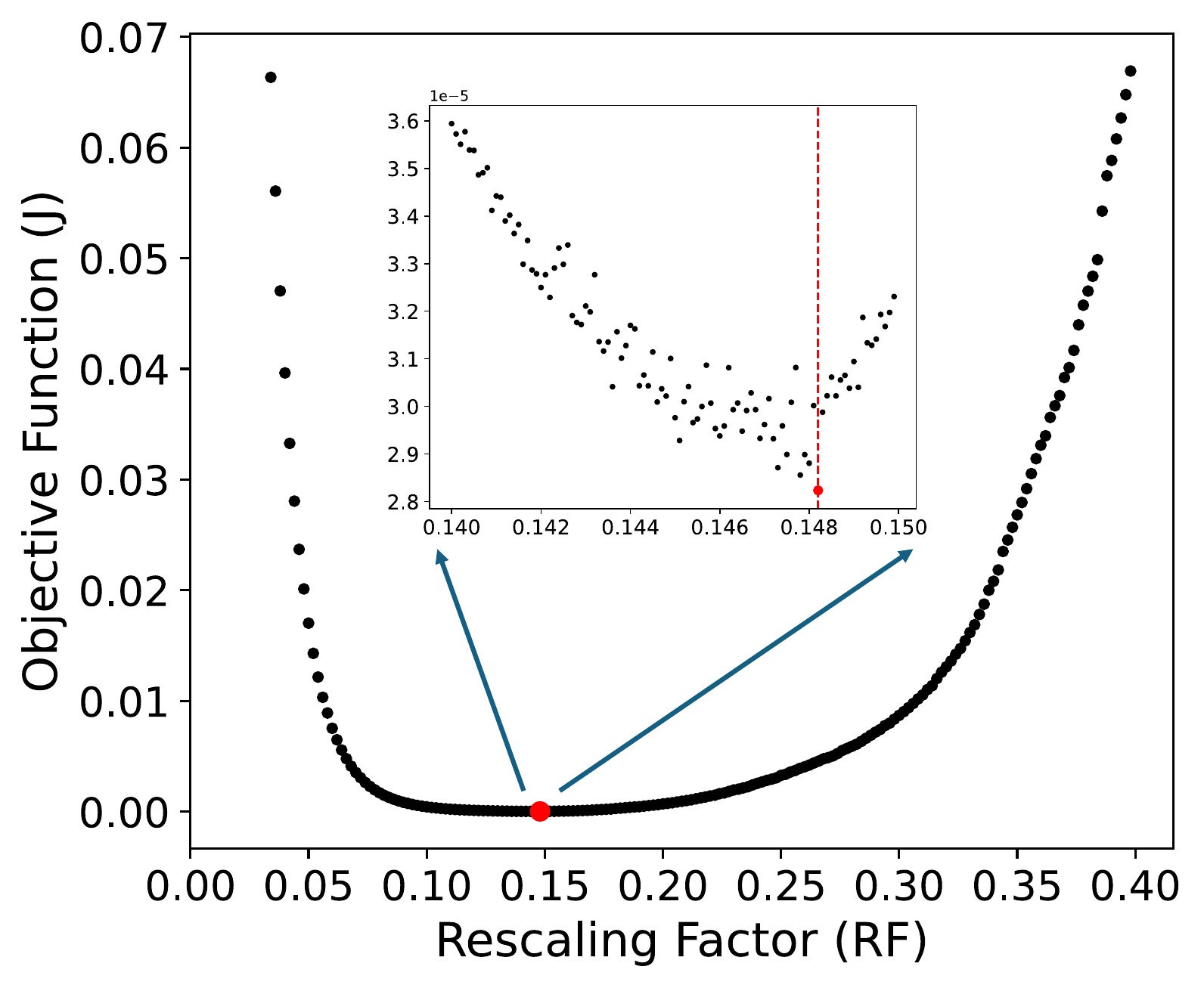}
    \caption{Variation of the objective function $J$ with the distance between the plasma boundary and CWS (RF) for the CWS obtained by extending the plasma boundary a distance RF. A scan is performed between 0.01 m and 0.4 m with a step value of $2\times 10^{-3}$ m. Then, a scan is performed between 0.14 m and 0.15 m with a step value of $1\times10^{-4}$ m. The value of RF obtained in this scan that minimizes the objective function is 0.1482 m.}
    \label{fig:sweeping_rescaled}
\end{figure}

\begin{figure}
    \centering
    \includegraphics[trim={0.5cm 0.6cm 0.8cm 1.0cm}, clip, width=0.9\linewidth]{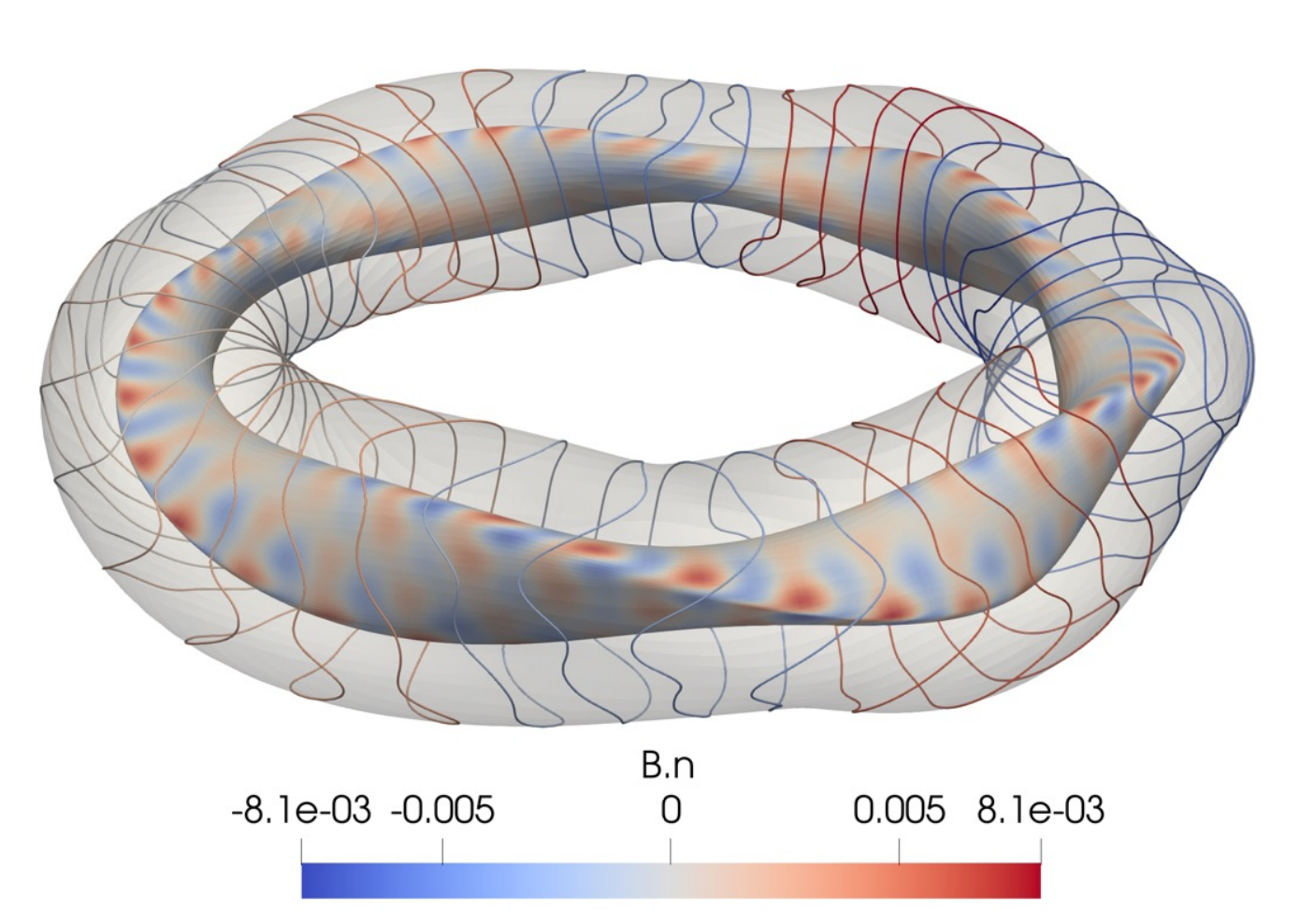}
    \caption{Set of coils obtained from an optimization targeting an approximately quasisymmetric magnetic field in a coil winding surface projected outward for a distance of 0.1482 m. Both the winding surface and the plasma have a major radius of 1 m.}
    \label{fig:rescaled}
\end{figure}

\begin{figure}
    \centering
    \includegraphics[trim={0.0cm 0.90cm 0.0cm 0.0cm}, clip, width=1\linewidth]{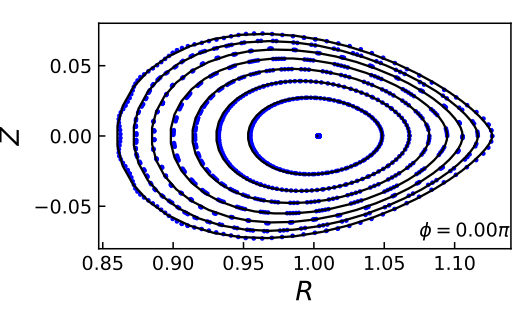}
    \caption{Superposition of magnetic surfaces of the target equilibrium at constant toroidal angle $\phi$ (black), and the Poincaré plot resulting from tracing magnetic field lines created by the optimized bounded coils (blue) in a CWS obtained as an offset of the plasma boundary, at the cylindrical angle $\phi = 0$.}
    \label{fig:rescaled_poincare}
\end{figure}

Finally, we optimize coils bound to a non-axisymmetric winding surface.
The CWS used here is similar to the benchmark cases used in previous CWS optimizations \cite{Paul2018}, namely, we rescale the plasma boundary outward by a fixed distance RF along the direction of the normal vector to the boundary. 
In this case, the rescaling factor corresponds to the distance between the CWS and the boundary surface at all points, and therefore to the coil-to-surface distance.
For this case, we again perform two scans to find the optimal coil-plasma distance, which are shown in Fig.~\ref{fig:sweeping_rescaled}. In the first scan, RF was varied from $0.01$ m to $0.40$ m with a step of $2\times 10^{-3}$ m, and in the second, from $0.14$ m to $0.15$ m with a step of $1\times10^{-4}$ m. The value of RF that minimizes the objective function is RF $ = 0.1482$ m.
Then, using this value for the rescaling factor, an optimization with 2000 iterations was done.
The resulting coil shapes are shown in Fig.~\ref{fig:rescaled}. The value obtained for the quadratic flux is $1.560\times 10^{-5}$ and the maximum field error on the surface is $\mathbf B \cdot \mathbf n = 8.1 \times 10^{-3}$.
A Poincaré plot at $\phi=0$ for the Biot-Savart resulting from the coils is shown in Fig.~\ref{fig:rescaled_poincare}.
Owing to the small $J$ and error fields, a very close agreement between the magnetic field lines (blue) and the target magnetic flux surfaces (black) is obtained throughout the whole volume.

\section{Conclusions}

In this work, a new method to design stellarator coils bound to a winding surface is developed. This allows coil simplicity metrics and winding surface metrics to be added to stellarator optimization studies independently. This combines the advantages of filamentary and winding surface optimizations, making highly constrained designs, such as the ones based on coil deposition on a surface, more feasible.
This approach is then used to optimize coils on two different winding surfaces, one resulting from the rescaling of the plasma boundary surface, and the other being a circular torus. The plasma boundary used is an approximately quasisymmetric one with five field periods. It is found that both approaches allow reproducing the boundary and inner set of nested flux surfaces of the target field.
The winding surface is optimized by performing a scan to obtain the optimal coil-to-surface distance. Both winding surfaces yield similar optima, at approximately 0.15 m from the plasma boundary for a stellarator with a major radius of 1 m and an aspect ratio of 10. The objective function $J$ is seen to have a parabolic dependence on the coil-plasma distance.
We note that while the coils obtained here are able to reproduce the target field, their maximum curvature and mean squared curvature are higher than previous coil designs for similar equilibria \cite{Wechsung2022,Jorge2023,Jorge2024}.
This is likely due to the highly constrained space of possible coils. However, while coil curvature is increased, it is conceivable that novel manufacturing methods such as direct deposition of high-temperature superconducting coils on a surface might allow more precise coils to be engineered at high curvatures.

Another future work to be done includes looking at other magnetic field equilibria, and optimizing the coil winding surface simultaneously with the optimization of curves parameterized to it. Having non-fixed CWS could be beneficial by bringing more freedom for the shape of the coils, meaning more adaptability to the desired needs for achieving the best possible magnetic field.
This could be allied with single-stage optimization to optimize the plasma boundary, the coil winding surface, and the coils simultaneously, leading to a more complete tool in the design of the next generation of fusion reactors.

\section*{Acknowledgements}

We thank M. Landreman, E. Koleman, D. Panici and R. Conlin for useful discussions.
R. J. is supported by the Portuguese FCT-Fundação para a Ciência e Tecnologia, under Grant 2021.02213.CEECIND and DOI  \href{https://doi.org/10.54499/2021.02213.CEECIND/CP1651/CT0004}{10.54499/2021.02213.CEECIND/CP1651/CT0004}.
R. J was also supported by the National Science Foundation under Grant No. 2409066.
Simulations were carried out using the EUROfusion Marconi supercomputer facility.
This work has been carried out within the framework of the EUROfusion Consortium, funded by the European Union via the Euratom Research and Training Programme (Grant Agreement No 101052200 - EUROfusion). Views and opinions expressed are however those of the author(s) only and do not necessarily reflect those of the European Union or the European Commission. Neither the European Union nor the European Commission can be held responsible for them.
IPFN activities were supported by FCT - Fundação para a Ciência e Tecnologia, I.P. by project reference UIDB/50010/2020 and DOI  \href{https://doi.org/10.54499/UIDB/50010/2020}{10.54499/UIDB/50010/2020}, by project reference UIDP/50010/2020 and DOI \href{https://doi.org/10.54499/UIDP/50010/2020}{10.54499/UIDP/50010/2020} and by project reference LA/P/0061/202 and  DOI \href{https://doi.org/10.54499/LA/P/0061/2020}{10.54499/LA/P/0061/2020}.
This work used the Jetstream2 at Indiana University through allocation PHY240054 from the Advanced Cyberinfrastructure Coordination Ecosystem: Services \& Support (ACCESS) program which is supported by National Science Foundation grants \#213859, \#2138286, \#2138307, \#2137603 and \#2138296.
This research used resources of the National Energy Research
Scientific Computing Center, a DOE Office of Science User Facility
supported by the Office of Science of the U.S. Department of Energy
under Contract No. DE-AC02-05CH11231 using NERSC award
NERSC DDR-ERCAP0030134.
This research used resources of the Oak Ridge Leadership Computing Facility at the Oak Ridge National Laboratory, which is supported by the Office of Science of the U.S. Department of Energy under Contract No. DE-AC05-00OR22725.


\section*{Data Availability Statement}

The data that support the findings of this study are openly available in Zenodo at \href{https://doi.org/10.5281/zenodo.13207692}{https://doi.org/10.5281/zenodo.13207692}.

\bibliographystyle{unsrt}
\bibliography{references}

\end{document}